\documentstyle[aps,prb,floats,epsf]{revtex}

\begin{document}

\title{\normalsize\bf
ACTIVATED BEHAVIOR OF THE 0.7 2{\boldmath${e^2}/{h}$}
CONDUCTANCE ANOMALY IN QUANTUM POINT CONTACTS
}

\author{\normalsize
A.\ Kristensen$^{\dagger}$, H.\ Bruus$^{\dagger}$,
A.\ Forchel$^{\ddagger}$, J.B.\ Jensen$^{\dagger}$,
P.E.\ Lindelof$^{\dagger}$,\\ 
M.\ Michel$^{\ddagger}$, J.\ Nyg\aa rd$^{\dagger}$,
and C.B.\ S\o rensen$^{\dagger}$, 
}

\address{\normalsize\em
$^{\dagger}$Niels Bohr Institute fAPG,
Universitetsparken 5, DK-2100 Copenhagen\\
$^{\ddagger}$ Technische Physik, Universit\"at W\"urzburg,
Am Hubland, D-97074 W\"urzburg\\
{\small Contributed paper for ICPS24, Jerusalem, August 2-7, 1998}
}

\maketitle
\centerline{\begin{minipage}{106.5mm} {\small \vspace*{5mm}
The 0.7 conductance anomaly in the quantized
conductance of trench etched GaAs quantum point contacts is studied
experimentally. The temperature dependence of the anomaly measured
with vanishing source-drain bias reveals the same activated behavior
as reported earlier for top-gated structures. Our main result is
that the zero bias, high temperature 0.7 anomaly found in activation
measurements and the finite bias, low temperature 0.9 anomaly found 
in transport spectroscopy have the same origin: a density dependent
excitation gap.}
\end{minipage}}

\vspace*{8mm} 
\noindent
The quantized conduction $G$ through a narrow point contact is one of
the key effects in mesoscopic physics, and deviations from perfect
quantization, such as the so-called 0.7 structure\cite{Thomas96}
appearing around $0.7$ times the conductance quantum $2 e^{2}/h$, are
important to understand. Firm conclusions regarding the origin of the
0.7 structure have been difficult to obtain partly due to the narrow
temperature range (0.1-4~K) in which the effect can be studied in
conventional split gate GaAs quantum point contacts, where relatively
close lying one-dimensional subbands are formed. Further progress has been
provided with the appearance of strongly confined GaAs quantum point
contacts using a combination of shallow etching and a top
gate.\cite{ak98a} These devices, roughly 50~nm wide and
100~nm long, have a subband energy spacing up to 20~meV and exhibit
conduction quantization up to around 30~K. They were later used to study
the temperature dependence deviations from perfect conductance 
quantization.\cite{ak98b} At low temperature ($T\approx 0.05$~K) almost
ideal quantized conductance is observed for the first conduction
plateau, but significant deviations develop at higher temperature
($T>1$~K). The 
enlarged temperature range allowed for the observation of activated
temperature dependence of these deviations: $\delta G(T) \propto
\exp(-T_a/T)$. Furthermore, by changing the top gate voltage it was
found that $T_a$ increases with increasing density. 

\begin{figure}[t]
\centerline{\epsfysize=55mm\epsfbox{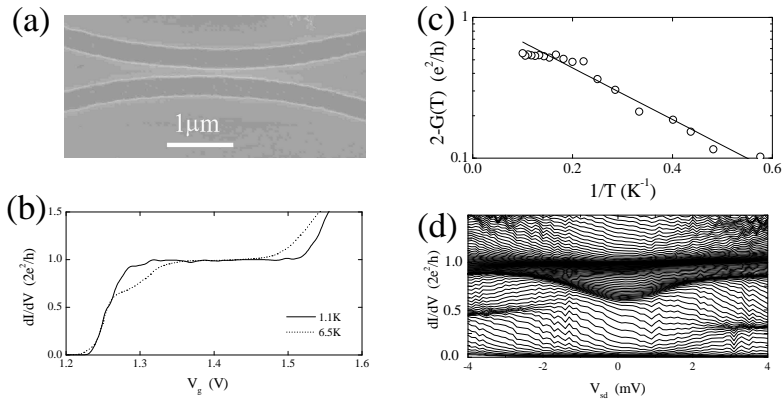}}
\caption{\label{fig:geometry}
Measurements on the $R = 10$~$\mu$m trench etched device. (a)
SEM-picture of the point contact showing the semicircular
trenches. (b) The first conductance plateau at $T=1.1$~K and
6.5~K. (c) Deviation from perfect quantized conductance
for $V_{\rm g}=1.3$~V; the slope
yields the activation energy $T_a=4.2$~K. (d) $dI/dV$ {\it vs.\/}
$V_{\rm sd}$ for $V_{\rm g}$ in the range 1.2-1.6~V. Conductance
plateaus appear as condensing of the curves.
}
\end{figure}

\begin{figure}[t]
\makebox[0.49\textwidth][r]{\epsfysize=35mm \epsfbox{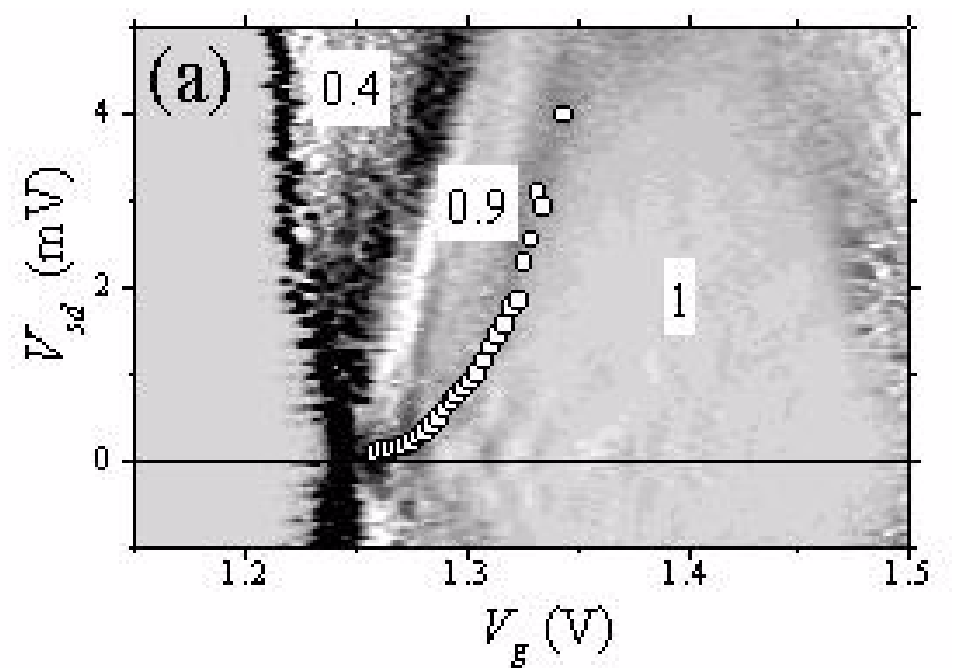}}
\makebox[0.49\textwidth][l]{\epsfysize=35mm \epsfbox{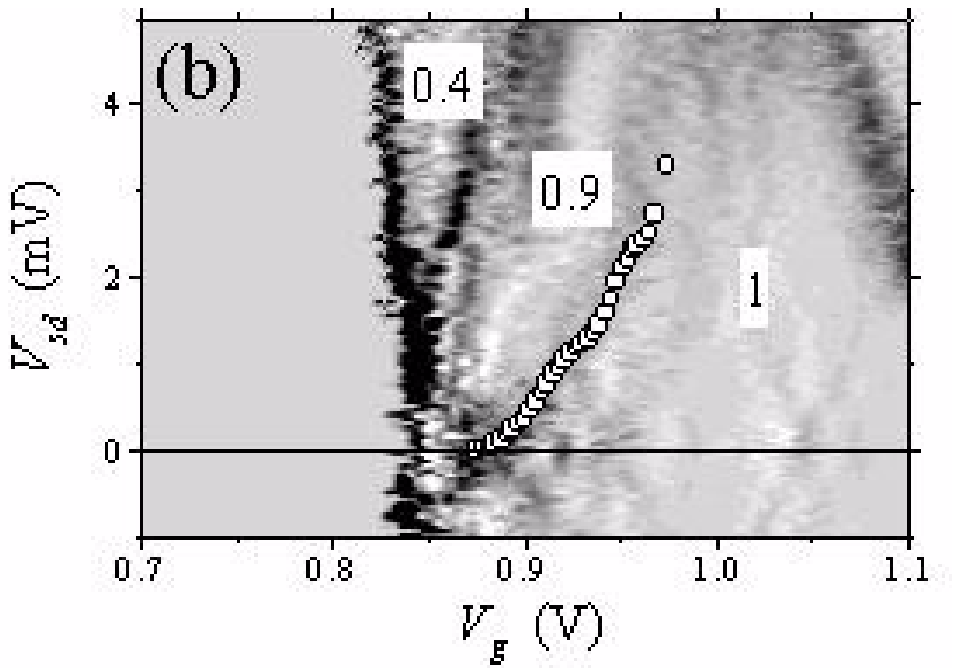}}
\makebox[0.49\textwidth][r]{\epsfysize=35mm \epsfbox{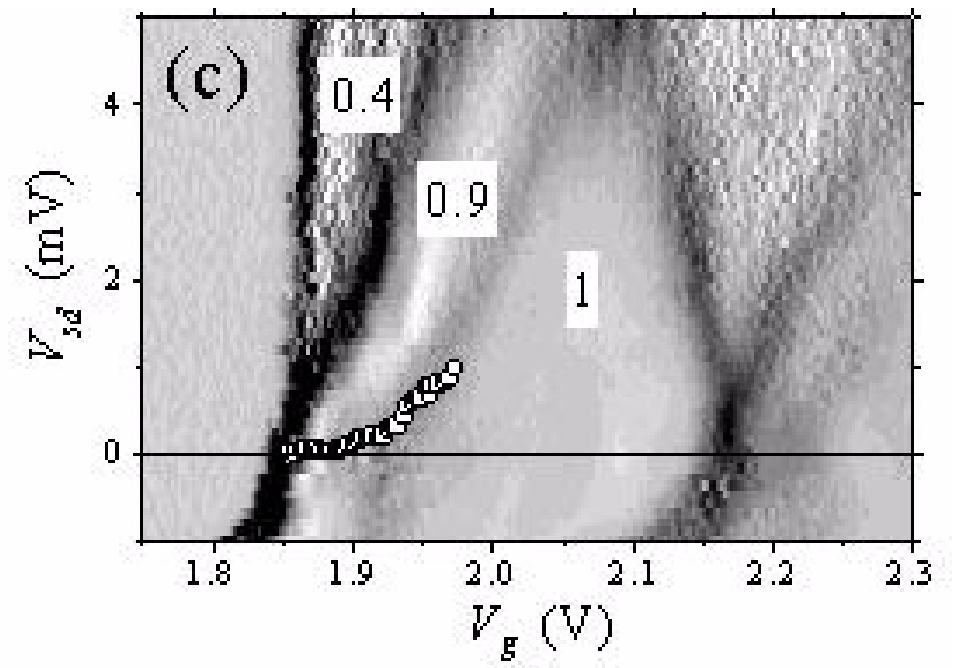}} \hfill
\makebox[0.49\textwidth][l]{\epsfysize=35mm \epsfbox{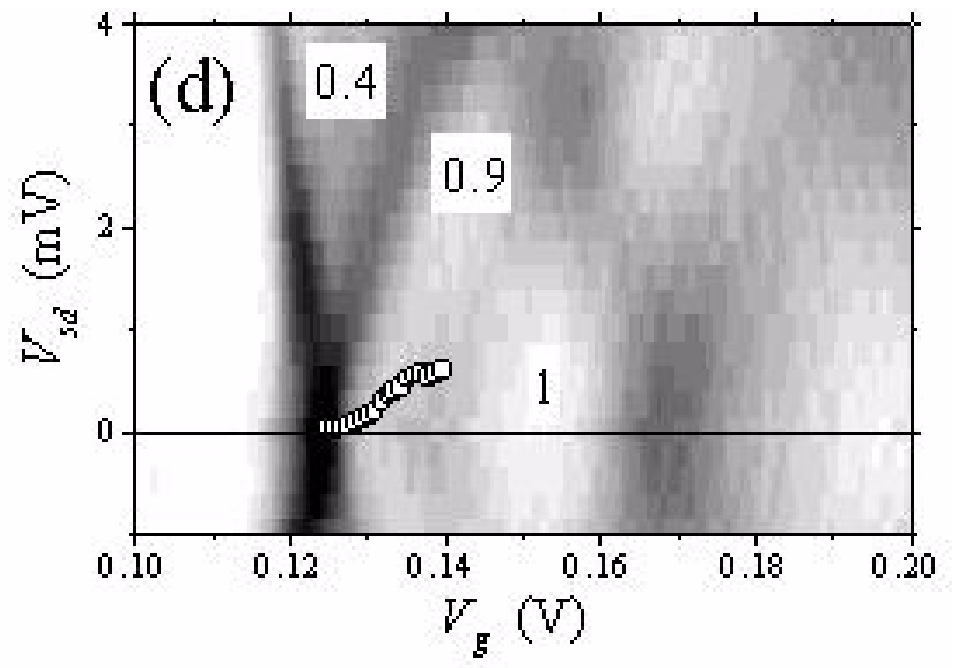}}
\caption{\label{fig:VgVsd}
A grayscale plot of $dG/dV_{\rm g}$ {\it vs.\/} $V_{\rm g}$
and $V_{\rm sd}$. Light regions correspond to conductance plateaus
while darker lines correspond transitions between plateaus. The white
circles are the activation energies $2 k_B T_a$ measured at zero bias.
(a)-(c) are the trench etched samples with $R=10$, 5, and 2~$\mu$m,
respectively. (d) is the top-gated sample from 
Ref.~3.
}
\end{figure}

In this paper we study the 0.7 conductance anomaly in trench etched
GaAs quantum point contacts. The GaAs heterostructure is the same as
in our previous work.\cite{ak98a,ak98b} However, we now fabricate the 
point contact by shallow etching 
semicircular shaped trenches 60~nm deep and
250~nm wide using $e$-beam lithography, see Fig.~\ref{fig:geometry}a. 
Three regions of the 2DEG are formed
electrically isolated from each other by the trenches. One region lies
between the semicircles and constitutes the point contact connecting
the source and drain reservoirs. The constriction is 
biased by a voltage denoted $V_{\rm sd}$. The other two regions lie 
inside the semicircles, and they can be used as side-gates by applying 
a gate voltage denoted $V_{\rm g}$. 
For devices with an etched width less than approximately 350~nm
the constriction is depleted of electrons at zero gate voltage. By
applying a positive gate voltage electrons are pulled towards the
constriction and for $V_{\rm g}$ larger than a critical pinch-off
voltage ($\sim\!1$~V) the point contact opens for electrical
conduction. The length of the constriction is defined through the
radius $R$ of the semicircles. We present three devices with 
$R=$ 2, 5, and 10~$\mu$m having the widths 150, 140, and 110~nm, 
respectively. The differential conductance $dI/dV$ is measured by
standard lock-in technique using an excitation voltage of 
6.3~$\mu$V rms at 117~Hz. Superposing the excitation voltage with 
a finite $V_{\rm sd}$ transport spectroscopy \cite{Patel91} can be
used to estimate the lowest 1D subband spacing, and we found it to
be 7.5, 8.0, and 9.5~meV for the three devices, respectively. In
Fig.~\ref{fig:geometry} two important results appear. (1) In 
Fig.~\ref{fig:geometry}b it is seen that the conductance anomaly is
present at $G\simeq0.7$ at the relatively high temperature 6.5~K but
absent at 1.1~K. As in our previous work on top-gated
structures\cite{ak98b} we find here that in the range 1-10~K the
conductance anomaly exhibits an activated behavior (see 
Fig.~\ref{fig:geometry}c) with a density dependent activation
temperature $T_a$. For the three devices $T_a$ ranges from 0 up to
0.5, 1.5, and 1.7~meV, respectively, as depicted in
Fig.~\ref{fig:VgVsd}. (2) In Fig.~\ref{fig:geometry}d it is seen how 
as a function of $V_{\rm sd}$ the conductance anomaly evolves smoothly
from 0.7 to 0.9 in units of $2e^2/h$.

We ascribe the conductance anomaly to a resonance with a bosonic 
excitation of energy $\varepsilon_b$ localized in the 
constriction.\cite{ak98b,Bruus98a} At zero bias and finite temperature 
the 0.7 structure arises from scattering against the thermally excited 
bosons. The activation temperature $T_a$ is therefore given by 
$k_{\rm B} T_a = \varepsilon_b$. The conductance anomaly is
investigated further using transport spectroscopy at low temperature, 
0.3~K. In such measurements resonances due to the boson excitation 
gap $\varepsilon_b$ at the Fermi level ought to show up.
We predict the position of these resonances by a simple model where
$V_{\rm sd} = \mu_1-\mu_2$ and the effective Fermi level in the middle 
of the constriction is defined by $(\mu_1+\mu_2)/2$, $\mu_1$ and $\mu_2$
being the chemical potential of the left and right contacts, respectively. 
Resonance is then obtained for $V_{\rm sd}^{\rm res}/2 = \varepsilon_b$,
and the predicted resonance condition is therefore $V_{\rm sd}^{\rm res} = 
2 k_{\rm B} T_a$. The result is shown in Fig.~\ref{fig:VgVsd} in the 
form of grayscale plots. The light diamond shaped regions correspond 
to flat plateau regions while the dark lines are transitions between them.
This overall structure is in accordance with previous reported 
results\cite{Thomas98}. The new and major result of our work is
that the measured values of $2 k_B T_a$ lies right on top of the
gray transition line between the 1.0 plateau region and the 0.9 
anomalous region. This strongly suggests a common origin of the zero
bias, high temperature 0.7 anomaly and the finite bias, low temperature
0.9 anomaly.

The change in conductance from 0.7 at zero bias to 0.9 at finite bias
can be understood heuristically as follows. The transmission at zero 
bias is given by ${\cal T} = 1 - R_b$, where $R_b$ is the reflection
due to relaxation of a thermally excited boson. Since we measure
${\cal T}=0.7$ we have $R_b=0.3$. At zero temperature no bosons
exist, and scattering can only occur by exciting bosons with the
energy $\varepsilon_b$. At the characteristic finite bias $V_{\rm sd}
= 2 \varepsilon_b$ the electrons coming in at energies lower than
$\varepsilon_b$ can not be back scattered due to Pauli blocking of the
final states. Those between $\varepsilon$ and $2\varepsilon$ can be
back scattered with a reflection probability $R_b$. Adding the
contribution from these two groups we find ${\cal
T}=\frac{1}{2}\!\times\!1 + \frac{1}{2} \!\times\! (1-R_b)$ or ${\cal
T} = 1 - \frac{1}{2} R_b$. From $R_b=0.3$ follows ${\cal T}=0.85$. 

In conclusion we have shown that also in trench etched quantum point
contacts there is a conductance anomaly very similar to what has been
reported for other types of samples. We have shown a good agreement
between the zero bias activation temperature and finite bias electron
spectroscopy resonances due to an energy gap. Experimentally we find
that this gap is density dependent rising from 0~meV at pinch-off to
roughly 2~meV at the middle of the first plateau depending of the
length of the constriction. There is a tendency to achieve higher
values of the gap for longer samples with a given width. 
Furthermore, the conduction anomaly disappears as a strong in-plane 
magnetic field is applied.\cite{Thomas96}
To explain the experimental results we propose that the anomaly is due 
to a density dependent excitation gap. Originally, we associated this gap
with a plasmon localized in the constriction.\cite{ak98b,Bruus98a} 
However to explain not only the density and temperature dependence, but 
also the magnetic field dependence\cite{Thomas96} and the length 
dependence reported here, we now tend to relate it to magnon
excitations of the possibly antiferromagnetically ordered quasi 1D
ground state\cite{Gruner94} of the constriction.

The work is supported by EU-LTR: QSWITCH 20960/30960,
A.K.\ by the Danish Technical Research Council Grant No.\
9701490, and  H.B.\ by the Danish Natural Science Research
Council Grant No.\ 9600548.


\begin{thebibliography}{99}

\vspace*{-5mm}
\bibitem{Thomas96} K.J. Thomas {\it et al.}, 
  {\em Phys. Rev. Lett.} {\bf 77}, 135 (1996). 
\bibitem{ak98a} A. Kristensen {\em et al.},
  {\em J. Appl. Phys.} {\bf 83}, 607 (1998), and
{\em Solid-State Electronics} {\bf 42}, 1103 (1998).
\bibitem{ak98b} A. Kristensen {\em et al.}, 
  {\em Physica} B (at press, 1998), cond-mat/9807277.
\bibitem{Patel91} N.K. Patel  {\it et al.}, 
  {\em Phys. Rev.} B {\bf 44}, 13549 (1991). 
\bibitem{Bruus98a} H. Bruus and K. Flensberg, 
  {\em Semicond. Sci. Technol.} {\bf 13}, A30 (1998).
\bibitem{Thomas98} K.J. Thomas {\it et al.}, 
  {\em Phys. Rev.} B (at press, 1998), cond-mat/9806242 .
\bibitem{Gruner94} G. Gr\"{u}ner, 
  {\em Rev. Mod. Phys.} {\bf 66}, 1 (1994).

\end{thebibliography}
\end{document}